\documentclass[11pt]{article}
\usepackage[letterpaper,margin=2cm,includehead,nomarginpar,headheight=2cm]{geometry}
\usepackage{amsmath,amssymb,fancyhdr,graphicx,xcolor}
\usepackage{newpxtext,newpxmath,hyperref,sectsty,lastpage,soul}
\hypersetup{breaklinks=true, bookmarks=true, pdfauthor={Valentin Vergara Hidd}, colorlinks=true, citecolor=gmu-green, urlcolor=gmu-green, linkcolor=gmu-coral, pdfborder={0 0 0}}
\definecolor{gmu-green}{RGB}{30,98,56}
\definecolor{gmu-gold}{RGB}{226,168,43}
\definecolor{gmu-coral}{RGB}{243,112,33}

\usepackage{lastpage}

\title{\sc The Rhythms of Transient Relationships: Allocating time between weekdays and weekends}
\author{Valent\'{i}n Vergara Hidd\thanks{George Mason University. \href{mailto:vvergara@gmu.edu}{vvergara@gmu.edu}} \and
  Mailun Zhang\thanks{George Mason University. \href{mailto:mzhang23@gmu.edu}{mzhang23@gmu.edu}} \and
  Simone Centellegher\thanks{Fondazione Bruno Kessler. \href{mailto:centellegher@fbk.eu}{centellegher@fbk.eu}} \and
  Sam Roberts\thanks{Liverpool Johns Moores University. \href{mailto:S.G.Roberts1@ljmu.ac.uk}{S.G.Roberts1@ljmu.ac.uk}} \and
  Bruno Lepri\thanks{Fondazione Bruno Kessler. \href{mailto:lepri@fbk.eu}{lepri@fbk.eu}} \and
  Eduardo L\'{o}pez\thanks{George Mason University. \href{mailto:elopez22@gmu.edu}{elopez22@gmu.edu}}
}
\date{May, 2023}

\begin{document}
\maketitle

\begin{abstract}
A fundamental question of any new relationship is, will it last? Transient relationships, recently defined by the authors, are an ideal type of social tie to explore this question: these relationships are characterized by distinguishable starting and ending temporal points, linking the question of tie longevity to relationship finite lifetime. In this study, we use mobile phone data sets from the UK and Italy to analyze the weekly allocation of time invested in maintaining transient relationships. We find that more relationships are created during weekdays, with a greater proportion of them receiving more contact during these days of the week in the long term. The smaller group of relationships that receive more phone calls during the weekend tend to remain active for more time. We uncover a sorting process by which some ties are moved from weekdays to weekends and \textit{vice versa}, mostly in the first half of the relationship. This process also carries more information about the ultimate lifetime of a tie than the part of the week when the relationship started, which suggests an early evaluation period that leads to a decision on how to allocate time to different types of transient ties.
\end{abstract}

\newpage

\section{Introduction}
Social relationships are an important part of people's everyday life. The number and properties of these social relationships have large effects on people's physical and mental health~\cite{hawkley10_lonel_matter,holt-lunstad10_social_relat_mortal_risk,pinquart01_influen_lonel_older_adult,williams18_inter_emotion_regul,holt-lunstad15_lonel_social_isolat_as_risk_factor_mortal}. Although much of the literature in this field focuses on long-lasting and subjectively important relationships (such as family or very long term friends)~\cite{burt00_decay_funct,hogan07_visual_person_networ,roberts10_commun_social_networ}, in reality a considerable portion of people's communication is dedicated to relatively short-term contacts (lasting in the order of a few months, perhaps a year). To begin to unravel this overlooked portion of people's communication, recent research has defined and described transient social relationships~\cite{hidd22_stabil_trans_relat}: they are short-lived ties with an identifiable beginning and where communication gets suspended for a considerable period of time that suggests the end of the most active part of relationship. Even though they have an ephemeral nature, they account for a significant proportion of communication activity: even if these relationships remain active for 5 or 6 months, they roughly receive between 1 to 3 phone calls every 15 days, which signals their importance~\cite{hidd22_stabil_trans_relat}. Among the outstanding questions to be tackled, it is still unclear what contributes to transient relationships emergence and what factors affect their duration. Crucially, in light of the large percentages of communication that transient relationships receive (ranging between $17\%$ and $45\%$ depending on the cohort~\cite{hidd22_stabil_trans_relat}), the dynamics of ego networks cannot be understood without taking them into account.

Of the few things known, previous research identifies a positive relationship between the amount of time a transient tie remains active (lifetime) and call volume~\cite{Navarro2017,hidd22_stabil_trans_relat}. However, it is also known that time is an inelastic resource~\cite{miritello13_time_as_limit_resour}, and therefore people need to make multiple decisions that enable meaningful social relationships to be maintained, so their ``need to belong'' (to have satisfying and meaningful connections with others) is met \cite{baumeister95_need_to_belon}, and less attention is given to those relationships perceived as less important. This raises the natural question of allocation of call timing, which may facilitate interaction with the variety of social ties people maintain with others.

Regarding the allocation of timing dedicated to social interactions, there is a large body of research examining circadian rhythms - how people's activities differ during the time of the day~\cite{panda02_circad_rhyth_from_flies_to_human,monsivais17_track_urban_human_activ_from,castaldo21_rhyth_night}, and particularly their communication patterns~\cite{aledavood15_daily_rhyth_mobil_telep_commun,aledavood18_social_networ_differ_chron_ident,aledavood22_quant_daily_rhyth_with_non,bhattacharya18_social_physic}. However, much less is known about how people allocate their time between their contacts considering the day of the week, or more broadly, weekday and weekend time. This information might provide insights into the association between the allocation of communication time and the nature of social relationships.

Thus, in order to advance our understanding of transient relationships in relation to the timing allocation decision process, in this paper we analyze mobile phone call records for people in two distinct demographic groups, which contrast in the life stage and country of ego. These records are from a period of time in which the bulk of non-face-to-face communication between individuals was through mobile calls because more recent phone-based communication tools such as WhatsApp, Messanger, and others were not available (phones had not yet become ``smart'')~\cite{roberts2011,saramaki2014persistence,Centellegher2016}
 (see subsection~\ref{sec:data} below for details about the data, including their time frame). Using these data, we characterize the formation of new transient relationships and their lifetimes and study how these two features are associated with the timing of when they occur, focusing on the difference between weekdays and weekends. We find that, while more relationships start on weekdays, the ones with longer lifetimes either start on or migrate to Saturday and Sunday. This pattern is consistent with the desire to organize the dynamics of people's social networks in order to optimize their value to the ego. This effect is similar to the variations known as circadian rhythms of human communication, which also emerge to better serve each ego's communication needs, such as more frequent contact to emotionally closer ties during evening and nights~\cite{aledavood15_daily_rhyth_mobil_telep_commun}. The relationships that migrate also indicate consistency with a homophily-driven~\cite{McPherson,Asikainen,Kossinets} evaluation period in which new relationships are assessed based on certain features~\cite{dunbar2018anatomy}, and communication is directed toward them according to the results of this evaluation.

``Weekend effects'' (or at least differences between weekdays and weekends) have been observed in physical activity~\cite{sigmundova16_weekd_weeken_patter_physic_activ} and sleeping patterns~\cite{taillard21_sleep_timin_chron_social_jetlag,paine16_differ_circad_phase_weekd_sleep}. In both examples, the outcome is the result of people allocating their time differently at weekends compared to weekdays. This paper extends the analysis of these differences to patterns of communication between the ego and specific members of their social network. Particularly, we examine phone call activity between ego and \textit{transient} alters on weekdays and weekends, to determine whether the nature of social relationships is associated with the allocation of calling time.

\section{Materials and Methods}
\label{sec:methods}
\subsection{Basic Definitions and Data}
\label{sec:data}
In this paper, we use mobile phone call data from two demographic groups, one located in the United Kingdom and the other in Italy which has been fully described in previous publications \cite{roberts2011, Centellegher2016}. The data for each country are constituted by the set of phone communication records from study participants (egos) and all the people they communicate with (alters) over 18 months in the United Kingdom (UK) between March 2007 and August 2008, and 24 months in Italy (IT), between January 2013 and December 2014. As additional details for both countries, the UK data follows 30 secondary school students in their transition to university or the labor market and records their phone call activity across 546 days~\cite{roberts2011}. IT data contains phone calls from 142 parents with children aged 0 to 10~\cite{Centellegher2016} over 700 days.

Consider ego $i$ and all of its alters, denoted here by $x,y,\ldots.$ For ego $i$ and one of its alters $x$, we use $c_{ix} \in \{1,2,3,\ldots,n_{i,x}\}$ as a counter for all outgoing phone calls from $i$ to $x$. Each of these calls takes place on a particular day $t(c_{i,x})$ of the corresponding study. The first day of the study is $0$ and the last day is $T_{\mathcal{E}}$ where $\mathcal{E} =\{{\rm UK}, {\rm IT}\}$. This means that $0\leq t(c_{i,x})\leq T_{\mathcal{E}}$ for each and every pair $ix$. By matching these days with their corresponding day of the week in the respective study, we can analyze communication activity as it relates to weekdays and weekends. Thus, for an alter that remains active in ego's network for $\ell_{i,x}$ days, the relationships \textit{lifetime}, we can identify a first call classification between weekday and weekend depending on the day in which the first phone call took place, as well as a communication volume classification, in which we measure the part of the week at which the relationship is more active.

\subsection{Transient relationships selection}
\label{sec:transientselection}
In order to identify transient relationships, here we use the same criteria introduced in~\cite{hidd22_stabil_trans_relat} and briefly explain it for the sake of clarity.

The criteria we employ to label a relationship as transient involve several conditions. First, transient relationships in the UK cannot exceed a maximum lifetime $\ell$ of $\ell \leq \mathcal{L} = 270$. Second, the first observed phone call between ego and the transient alter occurs at least $t = t_{s} = 180$ days after the beginning of data collection. The reason for this parameter is that after $180$ days, most of the individuals in the UK data moved from their hometown for their first year at university, and virtually all of their social networks changed dramatically from this point forward~\cite{saramaki2014persistence}.

In order to utilize the Italian data, we must introduce one additional definition. The study that produced this data~\cite{Centellegher2016} was conducted with rolling recruitment which means that, in contrast to the UK, $t=0$ may not be the first day of some egos in the study. Thus, we introduce the auxiliary variable $\tau$ that refers to the \textit{internal clock} of each ego such that the first observation of the ego in the data is marked as $\tau=0$ for that ego. To address the identification of Italian transient relationships, we now introduce the following rules. The maximum lifetime allowed is $\ell \leq \mathcal{L} = 270$, consistent with the UK. The first observed phone call between ego $i$ and transient alter $x$ is at least at day $\tau^{i}_{s} = 50$, where $\tau^{i}$ is a counter of days starting at the first observed phone call by ego $i$.

The third filter for transient relationships is the last observed phone call from ego to alter. For both countries, the last observed contact between ego and alter occurred at most $t = t_{w} = 60$ days before the end of data collection, i.e. $t(n_{i,x}) \leq T_{\mathcal{E_{{\rm UK}}}} - t_{w}$ and $\tau(n_{i,x}) \leq T_{\mathcal{E_{{\rm IT}}}} - t_{w}$, in order to improve the likelihood that the last observed phone call is the actual end of the relationship~\cite{miritello2011dynamical}.

In order to filter out phone calls where there is likely to be no meaningful social content, we remove all commercial numbers from the UK data. Due to the encoding of the data, this is not possible for Italy, but our results below show consistency between both countries, highlighting that whatever effects said commercial numbers may have on the qualitative results of our study are minimal.

Analysis of these data sets in~\cite{hidd22_stabil_trans_relat} showed that the vast majority of transient relationships are not particularly affected by the filters because they mostly start after a significant amount of time has passed after each of the studies begin and they end considerably before the studies end. As an example, the average day of entry for the UK was 119 and for Italy 283. Thus, whilst some effects of the timing of measurement could be present in our results, such effects are negligible.

All these filters yield 707 transient relationships in the UK (40.77\% of the relationships in the data), and 14191 transient relationships in Italy (59.94\% of the total).

\subsection{Classification of alters by first call and communication volume}
The first call classification of an alter between weekday or weekend is straightforward: it is constructed on the basis of the day of the week in which the first phone call took place. Monday to Friday are considered \textit{weekdays}, and Saturday and Sunday are considered \textit{weekend}. To track this, we introduce the random variable $\mathcal{C}_{o}$ which takes the values
\begin{equation}
  \label{eq:1}
  \mathcal{C}_{o} = \begin{cases} {\rm weekday}, &  t(c_{ix} = 0) \in \{{\rm Monday}, {\rm Tuesday},\ldots, {\rm Friday}\}\\
          {\rm weekend}, & {\rm otherwise}\end{cases}
\end{equation}

For the communication volume classification ($\mathcal{C}_{f}$), we use the overall sum of the duration of calls to the alter. Concretely, we construct an average of the time spent talking to a contact during weekdays over the number of weekdays when they communicate. For ego-alter pair $ix$, we denote this quantity as $\mathcal{D}^{W}_{i,x}$. Similarly for weekend days, we construct $\mathcal{D}^{E}_{i,x}$. Then, the communication volume classification of a tie as a \textit{weekend} relationship applies if $\mathcal{D}^{E}_{i,x} \geq \mathcal{D}^{W}_{i,x}$. On the other hand, if $\mathcal{D}^{E}_{i,x} < \mathcal{D}^{W}_{i,x}$ we classify the tie $i,x$ as a \textit{weekday} relationship. To be fully explicit, we next provide the mathematical expressions that correspond to this classification, but the intuition is the aforementioned one, which can be easily implemented directly in an algorithm.

Let $d(c_{ix})$ be the duration of call $c_{ix}$ between ego $i$ and alter $x$. Also, as an aid, we introduce the function $s(\cdot)$ that returns the day of the week of calendar day $t$. The average call time spent on active weekdays for $ix$ during the $\ell_{ix}$ days of the relationship is given by,
\begin{equation}
  \label{eq:durationW}
  \mathcal{D}^{W}_{i,x} = \frac{\sum\limits^{n_{ix}}_{c_{ix} = 1} U_{W}[s(t(c_{ix}))]\times d(c_{ix})}{U_W[s(t(c_{ix}=1))]+\sum\limits^{n_{ix}}_{c_{ix} = 2}U_W[s(t(c_{ix}))]\times\Theta\left[t(c_{ix})-t(c_{ix}-1)\right]},
\end{equation}
where $U_{W}(\cdot)$ is an indicator function that takes the value of 1 if the argument is in the set \{Monday, Tuesday, Wednesday, Thursday, Friday\} and 0 otherwise, and $\Theta(\cdot)$ is also an indicator function that evaluates to 0 when the argument is 0, and evaluates to 1 if the argument is $>0$. The numerator in Eq.~\ref{eq:durationW} is the sum of all time spent on phone calls from ego $i$ to alter $x$ that took place on all weekdays throughout alter's lifetime, while the denominator is the sum of \textit{weekdays} in which $i$ communicated with alter $x$. The role of the $\Theta(\cdot)$ function in the denominator is to count only different days of activity, avoiding multiple counts for any day in which more than one phone call took place.

Similarly, we can obtain the average call time over active \textit{weekend} days with
\begin{equation}
  \label{eq:durationE}
  \mathcal{D}^{E}_{i,x} = \frac{\sum\limits^{n_{ix}}_{c_{ix} = 1} U_{E}[s(t(c_{ix}))]\times d(c_{ix})}{U_E[s(t(c_{ix}=1))]+\sum\limits^{n_{ix}}_{c_{ix} = 2}U_E[s(t(c_{ix}))]\times\Theta\left[t(c_{ix})-t(c_{ix}-1)\right]},
\end{equation}
where the indicator function $U_{E}(\cdot)$ takes the value of 1 when the argument is in the set \{Saturday, Sunday\}, and 0 otherwise. The function $\Theta(\cdot)$ is the same used in Eq.~\ref{eq:durationW}. Then, in order to classify an alter we use
\begin{equation}
  \label{eq:Cf}
  \mathcal{C}_{f} = \begin{cases} {\rm weekday}, & \mathcal{D}^{E}_{i,x} < \mathcal{D}^{W}_{i,x}\\ {\rm weekend}, & \mathcal{D}^{E}_{i,x} \geq \mathcal{D}^{W}_{i,x} \end{cases}
\end{equation}

\subsection{Mutual Information}\label{sec:MI}
The mutual information scores for both classifications ($I(\ell, \mathcal{C}_{o})$ and $I(\ell, \mathcal{C}_{f})$, see Sec.~\ref{sec:results}) measure the amount of information one random variable contains about the other. In general, for discrete random variables $\mathbf{X}$ and $\mathbf{Y}$, we define the mutual information between them as
\begin{equation}
  \label{eq:MIdef}
  I(\mathbf{X}, \mathbf{Y}) = \sum_{x\in\mathbf{X}, \; y\in\mathbf{Y}}{\rm Pr}(\mathbf{X} = x, \textbf{Y} = y)\log_{2}\left[\frac{{\rm Pr}(\mathbf{X} = x, \textbf{Y} = y)}{{\rm Pr}(\mathbf{X} = x){\rm Pr}(\mathbf{Y} = y)} \right],
\end{equation}
where ${\rm Pr}(\mathbf{X} = x, \textbf{Y} = y)$ is a joint probability; ${\rm Pr}(\mathbf{X} = x)$, and ${\rm Pr}(\mathbf{Y} = y)$ are the marginal probabilities to draw $x$ and $y$, respectively. Mutual information is measured in bits, and the larger the value, the more bits of information for one variable are contained in the other.

\subsection{Bootstrapping for statistical testing}\label{sec:bootstrapping}
Both $I(\ell, \mathcal{C}_{o})$ and $I(\ell, \mathcal{C}_{f})$ are obtained using the entire set of alters $\mathcal{A}_{\mathcal{E}}$. Each alter in $\mathcal{A}_{\mathcal{E}}$ provides a tuple with three values: lifetime, first call classification, and communication volume classification. In order to provide more samples to test the consistency of our results, we draw a sample of 500 times the size as $\mathcal{A}_{\mathcal{E}}$ with replacement. Each alter of $\mathcal{A}_{\mathcal{E}}$ preserves its tuple. For the new sample, we compute mutual information, using the symbols $B(\ell, \mathcal{C}_{o})$ and $B(\ell, \mathcal{C}_{f})$ to distinguish from the mutual information scores using the original data. Then, we repeat the process 5000 times to produce distributions for $B(\ell, \mathcal{C}_{o})$ and $B(\ell, \mathcal{C}_{f})$.

\section{Results}
\label{sec:results}
\subsection{Number of relationships based on classification of part of the week}
In order to develop a description of how egos make allocation decisions for relationships to days of the week, we first study the statistics of such relationships when classified by the day of the week in which they are observed to communicate for the first time.

\begin{figure}[!h]
  \centering
  \includegraphics[width=0.95\linewidth]{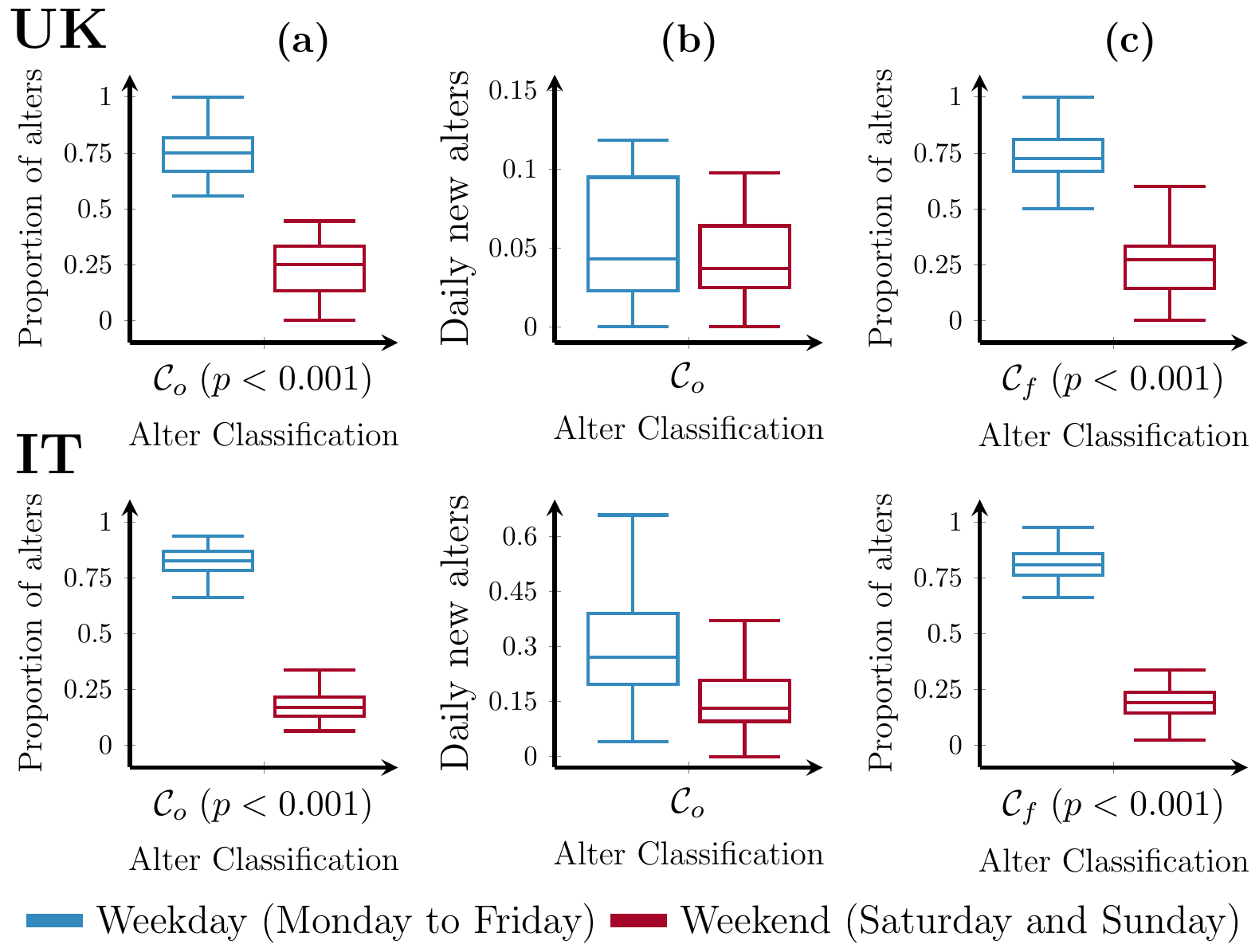}
  \caption{\textbf{Box plots for the proportion and the number of new daily transient alters as functions of their weekday or weekend classifications.}
    UK relationships are shown in the top row, and Italian relationships are at the bottom. Column \textbf{(a)} shows box plots for the proportion of alters ego by ego, based on the day of the first phone call. Column \textbf{(b)} shows the daily average new relationships at each part of the week, also based on $\mathcal{C}_{o}$. Column \textbf{(c)} shows the proportions of weekday and weekend alters based on the communication volume classification. For the number of relationships used in each group, see Table~\ref{tab:change-class}. Each plot shows also the $p$-value of a Mann-Whitney test between the weekday and weekend alters. All the plots show a greater amount of alters on weekdays in comparison to weekends.}
  \label{fig:WEstartend}
      \end{figure}

Figure~\ref{fig:WEstartend} column (a) presents the proportion of relationships for each group that start on a weekday (blue) or weekend (red), i.e. the ``first call classification'' ($\mathcal{C}_{o}$). We consider the proportion of relationships starting on a weekday/weekend for each ego and generate the corresponding box plot. As can be observed consistently among all countries and all groups, the proportion of relationships starting from Monday to Friday is significantly greater than those starting either on Saturday or Sunday. This is expected since one group contains more days than the other. To control for this difference, Fig.~\ref{fig:WEstartend} column (b) shows the daily average of new relationships, i.e. the total number of new relationships divided by the number of weekday/weekend days. Even though the tendency for transient ties is consistent with the results in column (a), i.e. more relationships start on weekdays, the effect is not as pronounced, and only statistically significant at the three star ($***$) level for the Italian data (only $**$ for the UK).

Our second statistical study of relationships, labeled as ``communication volume classification'' ($\mathcal{C}_{f}$), is done by classifying such relationships based on the duration of calls on weekdays against calls during the weekend (see Sec.~\ref{sec:methods}). At the tie level, we compare the average daily duration of phone calls placed from Monday to Friday, with the average daily duration of phone calls placed Saturday or Sunday. Each relationship is classified based on the type of day (weekday or weekend) in which there is more calling activity. Results are shown in Fig.~\ref{fig:WEstartend} column (c). Similar to $\mathcal{C}_{o}$, in both the UK and Italy data the proportion of weekday relationships is greater for the majority of egos.

In addition to the results above, we analyze each alter and its classifications in $\mathcal{C}_{o}$ and $\mathcal{C}_{f}$. This is an important and complementary result to those in Fig.~\ref{fig:WEstartend}, since for the majority of ties, their first call classification is where they will remain for the rest of their lifetimes, as shown by a $\chi^{2}$ test on the number of relationships for both countries. Particularly, the test shows that $\mathcal{C}_{o}$ is \textit{not} independent of $\mathcal{C}_{f}$ for the UK ($\chi^2 = 322.33; \; p < 0.001$), nor for Italy ($\chi^2 = 10723.33; \; p < 0.001$). Table~\ref{tab:change-class} shows the number of relationships that change classification from $\mathcal{C}_{o}$ to $\mathcal{C}_f$. By either using a subsample of only those relationships for which $\mathcal{C}_{o} = \mathcal{C}_{f}$, or the subset of alters for which $\mathcal{C}_{o} \neq \mathcal{C}_{f}$, results do not deviate from those presented in Fig.~\ref{fig:WEstartend}: i.e. a Mann-Whitney test shows that the median proportion of weekend transient relationships is less than the median proportion of weekend transient relationships ($p < 0.001$). This is true for $\mathcal{C}_{o}$ and $\mathcal{C}_{f}$ in both countries.

\begin{table}[!h]
  \centering
\caption{Number of transient relationships (and percentage of the total) conditional on their classifications.}
\label{tab:change-class}
\begin{tabular}{|ll|ccc|cccr|}
    \hline\hline
    & & \multicolumn{2}{c}{UK} & & & \multicolumn{2}{c}{IT} & \\
    & & \multicolumn{2}{c}{First Call Classification $\mathcal{C}_{o}$} & & & \multicolumn{2}{c}{First Call Classification $\mathcal{C}_{o}$} & \\
    & & Weekday & Weekend & & & Weekday & Weekend & \\
    \hline
    & Communication Volume & 486 & 34 & & & 11368 & 177& \\
    & Classification $\mathcal{C}_{f} = $ weekday & ($68.7\%$) & ($4.8\%$) & & & ($80.1\%$) & ($1.2\%$) & \\
    & & & & & & & &\\
    & Communication Volume & 52 & 135 & & & 373 & 2273 & \\
    & Classification $\mathcal{C}_{f} = $ weekend & ($7.4\%$) & ($19.1\%$) & & & ($2.6\%$) & ($16\%$) &\\
    \hline\hline
\end{tabular}
\end{table}

\subsection{Lifetime of a relationship}
Having established a difference between weekday and weekend relationships (consistent between $\mathcal{C}_{o}$ and $\mathcal{C}_{f}$), here we address their lifetime as an important feature even among the short-lived transient relationships. To measure how long any relationship remains active, we use the lifetime definition introduced in~\cite{hidd22_stabil_trans_relat}, $\ell_{i,x} = t(n_{i,x}) - t(c_{i,x} = 1)$.

\begin{figure}[!h]
  \centering
  \includegraphics[width=0.95\linewidth]{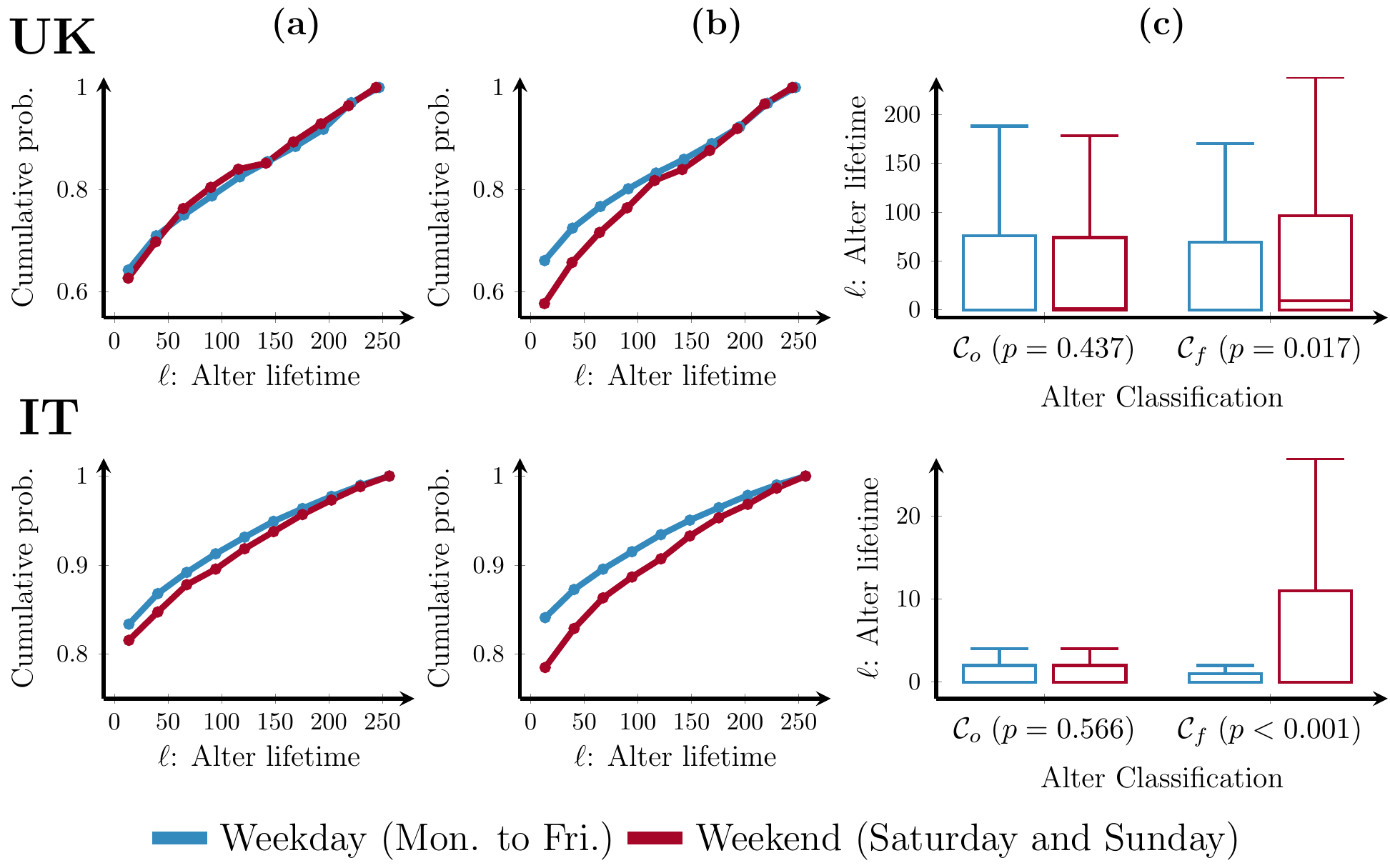}
 \caption{\textbf{Lifetime for weekday and weekend transient relationships, using $\mathcal{C}_{o}$ and $\mathcal{C}_{f}$.} Top row shows results for the UK, bottom row for Italy. Column \textbf{(a)} shows the cumulative distribution of lifetime conditional on first call classification. Column \textbf{(b)} shows the cumulative distribution of lifetime conditional on communication volume classification. Column \textbf{(c)} shows boxplots for lifetimes under both classifications, with $p$-values from a Mann-Whitney test for differences between weekday and weekend. Column \textbf{(b)} shows for both UK and IT the differences in lifetime distributions conditional $\mathcal{C}_{f}$. In particular, weekend relationships show a sharper increase with $\ell$, signaling the larger frequency of long lifetimes for relationships classified as weekend. These results are corroborated in Column \textbf{(c)} where the classification by $\mathcal{C}_{f}$ shows a clear increase of lifetimes for the weekends, compared to the same part of the week in $\mathcal{C}_{o}$.}
  \label{fig:WEell}
\end{figure}

Fig.~\ref{fig:WEell} column (a) shows the cumulative distribution of lifetime conditional to first call classification. Both weekday and weekend distributions are overlapped, which implies that no noticeable difference appears when separating relationships based on the day of the first call. In contrast, the cumulative distribution of lifetime conditional to communication volume classification in Fig.~\ref{fig:WEell} column (b) shows some separation between the curves, measuring a preference for longer lifetimes on weekends compared to weekdays. Also, the bulk of transient relationships tends to have an ultra-short lifetime (less than 60 days).

In order to further explore these differences between classifications, in Fig.~\ref{fig:WEell} column (c) we compare the boxplots of the previously described distributions. Here we confirm that for both countries, transient \textit{weekend} relationships (conditional to $\mathcal{C}_{f}$) have significantly longer lifetimes than their \textit{weekday} counterparts. Further, whilst the difference in lifetimes between weekday and weekend is not statistically significant for $\mathcal{C}_{o}$, it is for $\mathcal{C}_{f}$. This indicates that egos implement a sorting process in which long-lifetime relationships are moved to the weekend.

\subsection{Relationship Sorting}
Observed differences between $\mathcal{C}_{o}$ and $\mathcal{C}_{f}$ indicate egos sort some relationships away from their initial group. Although a tie may have a considerable amount of activity on weekends, there is no \textit{a priori} reason to believe such a relationship should have started on a weekend day. The reverse situation may also be true. Therefore, it is worth determining if there is a \textit{gradual} sorting process of relationships into $\mathcal{C}_{f}$, or if it happens more rapidly at the beginning of the relationship.

For both countries in this study, most relationships remain in the same classification as they started, i.e. relationships that started on a particular part of the week remain in the same classification throughout their lifetime (see Table~\ref{tab:change-class}). However, some relationships get shifted over time. For the relationships that do change from their initial classification, we want to consider differences in terms of \textit{when} in their lifetimes that change occurs. Let us define $a_{ix}$ as a daily counter for the elapsed duration of a relationship between ego $i$ and alter $x$, i.e. the number of days since the first observed phone call from ego $i$ to alter $x$. The normalized version of this quantity, $a_{ix}/\ell_{ix}$, allows for the comparison of alters with different lifetimes. Let us further define $a^{*}_{i,x}$ as the elapsed duration of the relationship at which point its classification by volume of communication is reached. We now study the cumulative distribution of $a^*_{ix}/\ell_{ix}$ across alters.

\begin{figure}[!h]
  \centering
  \includegraphics[width=0.95\linewidth]{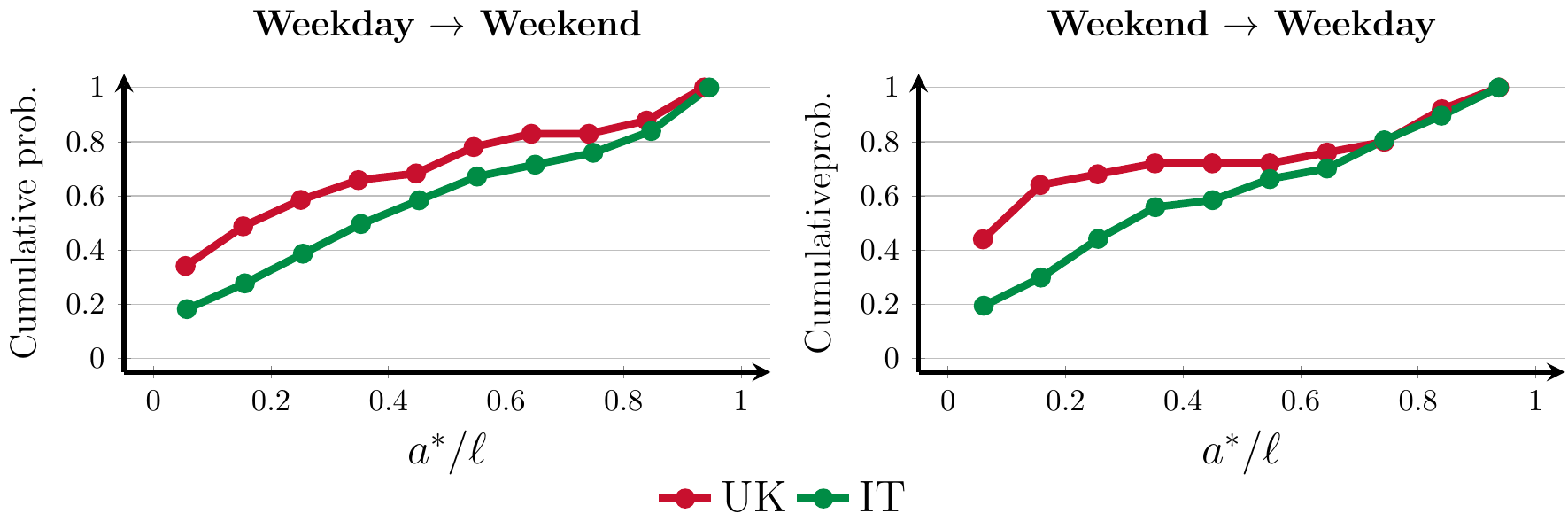}
\caption{\textbf{Cumulative distributions for the proportion of lifetime it takes a relationship to settle into communication volume classification.} In both plots, the horizontal axis represents the elapsed duration of a relationship when the communication volume classification is reached.}
  \label{fig:settling}
\end{figure}

Fig.~\ref{fig:settling} shows that approximately 60\% of transient relationships have settled into their final classification by half of their lifetime. This is true for relationships moving \textit{to} the weekend, and also for those moving \textit{from} the weekend. The former group seems to settle slightly faster than those that move from the weekend to weekdays. There is general consistency between countries, but alters in the UK reach their communication volume classification slightly faster, which is due to egos making more daily phone calls, on average than egos in Italy (UK daily average is $1.0092$; IT average is $0.4020$). Also for the UK, the fact that most participants are under a university calendar that encourages the formation of many new relationships is probably an additional contributor to the faster classification. All of this provides further evidence to support the idea of a sorting mechanism by ego, relatively early in the relationship.

\subsection{Weekend relationships are not randomly selected}
The results displayed in Fig.~\ref{fig:settling} offer a visual indication that ego sorts \textit{transient} relationships with longer lifetimes to the weekends. In order to test if the sorting of alters into the weekend category is statistically significant, we compute the mutual information between lifetime and first call classification $I(\ell, \mathcal{C}_{o})$ and lifetime and communication volume classification $I(\ell, \mathcal{C}_{f})$ and check if their values differ by an amount that cannot be explained by random chance.

To explain the intuition, if indeed there is a mechanism that sorts long lifetime alters into the \textit{weekend} group by the time of their communication volume classification (i.e. the end of their lifetime), we expect to find that $I(\ell, \mathcal{C}_{o}) < I(\ell, \mathcal{C}_{f})$. This would mean that classifying each ego-alter transient relation by the part of the week in which they communicate the most is more informative about the relationship's lifetime than the first day they communicate. Applying this method (details in Sec.~\ref{sec:MI}), we obtain the values for both countries shown in Table~\ref{tab:MI}. For both countries, we confirmed that $\mathcal{C}_{o}$ carries less information about lifetime than $\mathcal{C}_{f}$.

\begin{table}[!h]
  \centering
  \caption{Mutual information between lifetime and alter's first call classification ($I(\ell, \mathcal{C}_{o})$) and between lifetime and alter's communication volume classification ($I(\ell, \mathcal{C}_{f})$). Next to each value, we include the average over 5000 samples from a bootstrapping procedure using a sample size of 500 times the original data for each country. $\Delta B = \langle B(\ell, \mathcal{C}_{f}) \rangle  - \langle B(\ell, \mathcal{C}_{o}) \rangle$, a difference tested to be significantly different from 0 using an independent sample $T$-test, with $p$-values reported in the last column to the right. All mutual information values are in bits.}
  \label{tab:MI}
  \begin{tabular}{|lcccccc|}
    \hline \hline
    & M. I. & M. I. & M. I. & M. I. & & \\
    & Lifetime and & bootstrap & Lifetime and & bootstrap & $\Delta I^{b}$ & $p$-value\\
    & First Call & Average & Comm. Vol. & Average & & \\
    & $I(\ell, \mathcal{C}_{o})$ & $\langle B(\ell, \mathcal{C}_{o})\rangle$ & $I(\ell, \mathcal{C}_{f})$ & $\langle B(\ell, \mathcal{C}_{f})\rangle$ &  & \\
    \hline
    UK & $0.2249$ & $0.2250$ & $0.2361$ & $0.2362$ & $0.0112$ & $< 0.001$\\
    IT & $0.0153$ & $0.0153$ & $0.0171$ & $0.0171$ & $0.0018$ & $< 0.001$\\
    \hline\hline
  \end{tabular}
\end{table}

\begin{figure}[!h]
  \includegraphics[width=0.95\textwidth]{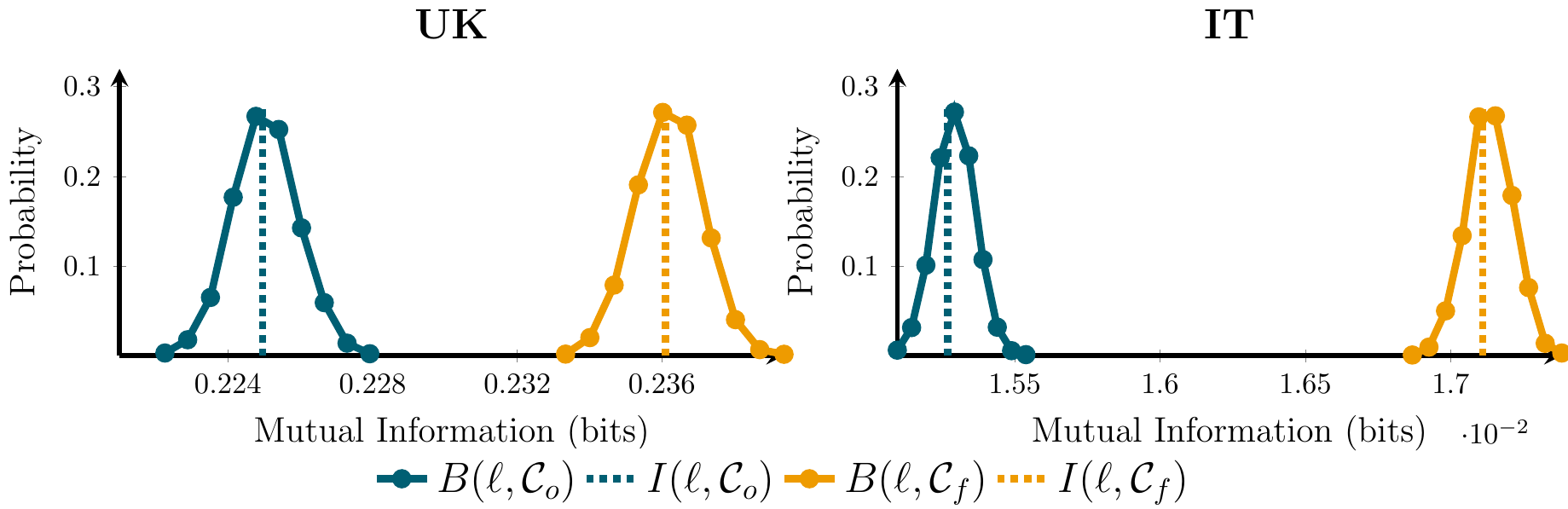}
  \caption{\textbf{Distribution of mutual information values from a bootstrap procedure using both classifications along with measured mutual information.} The left plot shows results from the UK, and the right plot from Italy. In both, the distribution $B(\ell, \mathcal{C}_{o})$ is shown in dark blue, and the corresponding distribution of $B(\ell, \mathcal{C}_{f})$ in yellow. All distributions were obtained using a bootstrapping method with a sample size of 500 times the original data, with 5000 repetitions. The corresponding measured values of mutual information are marked with dashed lines. Clearly, in both countries $I(\ell, \mathcal{C}_{o}) < I(\ell, \mathcal{C}_{f})$, and the bootstrap results show that the differences cannot be explained as the consequence of mere stochastic fluctuations.}
  \label{fig:MIboot}
\end{figure}

In order to perform our statistical test to determine if the difference between $I(\ell, \mathcal{C}_{o})$ and $I(\ell, \mathcal{C}_{f})$ found in Table~\ref{tab:MI} is significant, we created $5000$ samples for both mutual information scores using a bootstrapping method (details in Sec.~\ref{sec:bootstrapping}). Briefly, the method samples with replacement transient relationships and calculates for each sample a value of mutual information per classification, i.e. $B(\ell, \mathcal{C}_{o})$ and $B(\ell, \mathcal{C}_{f})$. The sample size in each bootstrap iteration is 500 times as the original data (e.g. for the $707$ transient relationships in the UK, $500\times 707 = 353,500$ randomly chosen relationships are taken to generate a single value of $B(\ell, \mathcal{C}_{o})$ and $B(\ell, \mathcal{C}_{f})$). The reason for the sample size is due to the distributions of $\ell$. Since longer lifetimes have a low probability of occurrence, sampling procedures tend to underrepresent long lifetimes with smaller sample sizes. The distributions for these $5000$ mutual information scores (in bits) are shown in Fig.~\ref{fig:MIboot}. No overlap is visible between the distributions of $B(\ell, \mathcal{C}_{o})$ and $B(\ell, \mathcal{C}_{f})$. We also used an independent sample $T$-test ($p < 0.001$ for both countries) to show that $\langle B(\ell, \mathcal{C}_{o})\rangle < \langle B(\ell, \mathcal{C}_{f})\rangle$. Additionally, we used a Kolgorov-Smirnov test to show that the distributions are in fact different ($p < 0.001$ for both countries).

The interpretation of the results above is that $\mathcal{C}_{o}$ carries less information about lifetime of alters than $\mathcal{C}_{f}$.

\section{Discussion}
In this study, we used mobile data from the UK and Italy to examine how the distribution of mobile communication over days of the week is related to the lifetime of transient alters.

One interesting finding is that it is more common to start a new relationship during a weekday, a surprising result considering that weekdays typically contain the bulk of work schedules, leaving less room for social activities. However, we should also consider that new transient relationships do not necessarily have to be friendship interactions. They can start as merely utilitarian contacts that are extinguished after their main purpose has been fulfilled (e.g. a group project at school, or a new client in a work environment). This result may be due to more opportunities to meet new alters across contexts, but our data cannot test this hypothesis. Regardless, we find consistency between the two countries analyzed, a result that suggests that more new relationships start during weekdays, i.e. Monday to Friday.

Then, we use the entire lifetime of a tie to construct the final classification based on the duration of calls, which in the majority of relationships corresponds to the initial day of contact. Overall, each ego has more relationships classified as \textit{weekday} ties than their \textit{weekend} counterparts. Thus, the existence of two distinct groups of alters of different sizes is coherent with the organization of our contacts in a layered structure~\cite{zhou2005organizational_social_groups,dunbar2018anatomy}, with the inner layer composed of the smallest group. In this theoretical framework of the ``circles of friendship'', the inner layer is not only distinguishable from the rest because of its size, but also concentrates the set of alters with more emotional closeness to ego~\cite{dunbar2018anatomy}. The patterns we find here suggest that transient relationships in which most communication activity occurs on weekends are closer to the inner (smaller) layers of egos' contact networks.

Further, we demonstrate that the lifetime of a transient alter is related to calling patterns, with alters that receive more of their communication at the weekend having longer lifetimes than those that receive more of their communication on weekdays. There is also evidence of a
``sorting'' process, where 60\% of transient relationships have reached their $\mathcal{C}_f$ value before 50\% of their lifetime. This suggests that egos selectively allocate their calling time on days of the week based on the characteristics of the alter, with \textit{weekend alters} having longer lifetime than \textit{weekday alters}.

Whilst currently there is a fragmentation of the channels of communication between ego and alter, now distributed among a plethora of instant message applications, social media sites, and email, our results are robust because at the time the data was collected, mobile calls were the main method of communication (``smart'' phones had not yet affected communication). However, a more critical question might be whether these new channels would indeed be able to affect our results? To this, we offer two reasons why we believe the actual social patterns seen in our study should be considered robust. First, there have been multiple studies that confirm that social patterns observed in phone calls resemble those found in emails~\cite{Godoy} and Facebook messages~\cite{DunbarArnaboldi}, lending support to the notion that new communication methods do not fundamentally change what people do socially. Second, we emphasize that the main contribution in terms of social and psychological value is the ego-alter relationship, not the channel used for communication. In this sense, communicating via the phone, short message service (SMS), and/or WhatsApp all occur to achieve the same objective. These considerations suggest that it is more likely that the effect of channel fragmentation is to complicate data collection of meaningful social patterns rather than lead to their fundamental change. Indeed, one could argue that this very situation makes our data particularly valuable by having been collected before it was too difficult to be collected again with little ambiguity or challenge.

All of these results have to be interpreted considering their limitations, mostly the uncertainty about the actual beginning and end of transient relationships. Even when including the filters described in section~\ref{sec:methods}, there is an inherent limitation to all phone call data, because it is not practical to obtain and analyze phone call records for more than a few years. In practice, this means that all samples are limited and cannot capture the complete set of alters for each ego. In this study, we considerably mitigate the risk of transient relationship classification error by our methods, particularly in the UK where at least the start of the relationship is selected in such a way as to virtually guarantee the alters are new due to the egos moving from school to University or the labor market. Furthermore, in~\cite{hidd22_stabil_trans_relat} it is shown that in fact, most relationships identified as transients are comfortably within the boundaries set by the transient relationship filters in both the UK and Italy, supporting the robustness of our results. Another limitation of our work is that it suggests an evaluation of relationships at their early stages, but does not explain the mechanism by which this evaluation takes place. Without a qualitative assessment of each alter, timestamps for phone calls limit the extent to which we can describe an evaluation mechanism.

In summary, there are qualitative differences between relationships maintained through communication mostly between Monday and Friday, and those with more communication during Saturday and Sunday. More new relationships are formed during the weekdays, but relationships with more communication on the weekend have a longer lifetime. This is an unexplored side of the large body of literature regarding calling patterns based on allocation of timing of relationships, with the only major example being circadian rhythms over \textit{all} relationships~\cite{aledavood15_daily_rhyth_mobil_telep_commun,aledavood18_social_networ_differ_chron_ident,aubourg20_novel_statis_approac_asses_persis,aledavood22_quant_daily_rhyth_with_non}. Additionally, these results are relevant in the discussion of the ``Seven Pillars of Friendship''~\cite{dunbar2018anatomy}, a homophily-driven theory that explains how friendships are formed. In this theory, social relationships are evaluated on a heuristic basis on seven factors of homophily in order to decide on their suitability as friendships. The heuristic is such that it allows for a ``rapid'' evaluation of a new relationship. The sorting process observed here supports the notion that an evaluation of every alter early in the relationship is taking place. Even if the time of the first encounter between ego and alter is (slightly more) random, the allocation of time by the end of the relationship is not. This explanation is also in line with the literature about the optimization of social interactions~\cite{tamarit18_cognit_resour_alloc_deter_organ_person_networ,tamarit22_beyon_dunbar_circl}. Broadly, the results presented here contribute to our understanding of the nature of relationships, even when the data is limited and does not include information on subjective evaluation of alters.

\bibliography{wEreferences}

\begin{thebibliography}{10}

\bibitem{aledavood22_quant_daily_rhyth_with_non}
Talayeh Aledavood, Ilkka Kivim{\"a}ki, Sune Lehmann, and Jari Saram{\"a}ki.
\newblock Quantifying daily rhythms with non-negative matrix factorization
  applied to mobile phone data.
\newblock {\em Scientific Reports}, 12(1):5544, 2022.

\bibitem{aledavood18_social_networ_differ_chron_ident}
Talayeh Aledavood, Sune Lehmann, and Jari Saram{\"a}ki.
\newblock Social network differences of chronotypes identified from mobile
  phone data.
\newblock {\em EPJ Data Science}, 7(1):46, 2018.

\bibitem{aledavood15_daily_rhyth_mobil_telep_commun}
Talayeh Aledavood, Eduardo L{\'o}pez, Sam G.~B. Roberts, Felix Reed-Tsochas,
  Esteban Moro, Robin I.~M. Dunbar, and Jari Saram{\"a}ki.
\newblock Daily rhythms in mobile telephone communication.
\newblock {\em PLOS ONE}, 10(9):e0138098, 2015.

\bibitem{Asikainen}
Aili Asikainen, Gerardo Iñiguez, Javier Ureña-Carrión, Kimmo Kaski, and
  Mikko Kivelä.
\newblock Cumulative effects of triadic closure and homophily in social
  networks.
\newblock {\em Science Advances}, 6(19):eaax7310, 2020.

\bibitem{aubourg20_novel_statis_approac_asses_persis}
Timoth{\'e}e Aubourg, Jacques Demongeot, and Nicolas Vuillerme.
\newblock Novel statistical approach for assessing the persistence of the
  circadian rhythms of social activity from telephone call detail records in
  older adults.
\newblock {\em Scientific Reports}, 10(1):21464, 2020.

\bibitem{baumeister95_need_to_belon}
Roy~F. Baumeister and Mark~R. Leary.
\newblock The need to belong: Desire for interpersonal attachments as a
  fundamental human motivation.
\newblock {\em Psychological Bulletin}, 117(3):497--529, 1995.

\bibitem{bhattacharya18_social_physic}
Kunal Bhattacharya and Kimmo Kaski.
\newblock Social physics: Uncovering human behaviour from communication.
\newblock {\em Advances in Physics: X}, 4(1):1527723, 2018.

\bibitem{burt00_decay_funct}
Ronald~S Burt.
\newblock Decay functions.
\newblock {\em Social Networks}, 22(1):1--28, 2000.

\bibitem{castaldo21_rhyth_night}
Maria Castaldo, Tommaso Venturini, Paolo Frasca, and Floriana Gargiulo.
\newblock The rhythms of the night: Increase in online night activity and
  emotional resilience during the spring 2020 covid-19 lockdown.
\newblock {\em EPJ Data Science}, 10(1):7, 2021.

\bibitem{Centellegher2016}
Simone Centellegher, Marco De~Nadai, Michele Caraviello, Chiara Leonardi,
  Michele Vescovi, Yusi Ramadian, Nuria Oliver, Fabio Pianesi, Alex Pentland,
  Fabrizio Antonelli, and Bruno Lepri.
\newblock The mobile territorial lab: a multilayered and dynamic view on
  parents' daily lives.
\newblock {\em EPJ Data Science}, 5(1):3, Feb 2016.

\bibitem{DunbarArnaboldi}
R.I.M. Dunbar, Valerio Arnaboldi, Marco Conti, and Andrea Passarella.
\newblock The structure of online social networks mirrors those in the offline
  world.
\newblock {\em Social Networks}, 43:39--47, 2015.

\bibitem{dunbar2018anatomy}
Robin I~M Dunbar.
\newblock The anatomy of friendship.
\newblock {\em Trends in cognitive sciences}, 22(1):32--51, 2018.

\bibitem{Godoy}
Antonia Godoy-Lorite, Roger Guimerà, and Marta Sales-Pardo.
\newblock Long-term evolution of email networks: Statistical regularities,
  predictability and stability of social behaviors.
\newblock {\em PLOS ONE}, 11(1):1--11, 01 2016.

\bibitem{hawkley10_lonel_matter}
Louise~C. Hawkley and John~T. Cacioppo.
\newblock Loneliness matters: a theoretical and empirical review of
  consequences and mechanisms.
\newblock {\em Annals of Behavioral Medicine}, 40(2):218--227, 2010.

\bibitem{hidd22_stabil_trans_relat}
Valent\'{i}n~Vergara Hidd, Eduardo L\'{o}pez, Simone Centellegher, Sam G.~B.
  Roberts, Bruno Lepri, and Robin I.~M. Dunbar.
\newblock The stability of transient relationships.
\newblock {\em Scientific Reports}, 13(6120), 2023.

\bibitem{hogan07_visual_person_networ}
Bernie Hogan, Juan~Antonio Carrasco, and Barry Wellman.
\newblock Visualizing personal networks: Working with participant-aided
  sociograms.
\newblock {\em Field Methods}, 19(2):116--144, 2007.

\bibitem{holt-lunstad15_lonel_social_isolat_as_risk_factor_mortal}
Julianne Holt-Lunstad, Timothy~B. Smith, Mark Baker, Tyler Harris, and David
  Stephenson.
\newblock Loneliness and social isolation as risk factors for mortality.
\newblock {\em Perspectives on Psychological Science}, 10(2):227--237, 2015.

\bibitem{holt-lunstad10_social_relat_mortal_risk}
Julianne Holt-Lunstad, Timothy~B. Smith, and J.~Bradley Layton.
\newblock Social relationships and mortality risk: a meta-analytic review.
\newblock {\em PLoS Medicine}, 7(7):e1000316, 2010.

\bibitem{Kossinets}
Gueorgi Kossinets and Duncan~J. Watts.
\newblock Origins of homophily in an evolving social network.
\newblock {\em American Journal of Sociology}, 115(2):405--450, 2009.

\bibitem{McPherson}
Miller McPherson, Lynn Smith-Lovin, and James~M Cook.
\newblock Birds of a feather: Homophily in social networks.
\newblock {\em Annual Review of Sociology}, 27(1):415--444, 2001.

\bibitem{miritello2011dynamical}
Giovanna Miritello, Esteban Moro, and Rub{\'e}n Lara.
\newblock Dynamical strength of social ties in information spreading.
\newblock {\em Physical Review E}, 83(4):045102, 2011.

\bibitem{miritello13_time_as_limit_resour}
Giovanna Miritello, Esteban Moro, Rub{\'e}n Lara, Roc{\'i}o
  Mart{\'i}nez-L{\'o}pez, John Belchamber, Sam~G.B. Roberts, and Robin~I.M.
  Dunbar.
\newblock Time as a limited resource: Communication strategy in mobile phone
  networks.
\newblock {\em Social Networks}, 35(1):89--95, 2013.

\bibitem{monsivais17_track_urban_human_activ_from}
Daniel Monsivais, Asim Ghosh, Kunal Bhattacharya, Robin I.~M. Dunbar, and Kimmo
  Kaski.
\newblock Tracking urban human activity from mobile phone calling patterns.
\newblock {\em PLOS Computational Biology}, 13(11):e1005824, 2017.

\bibitem{Navarro2017}
Henry Navarro, Giovanna Miritello, Arturo Canales, and Esteban Moro.
\newblock Temporal patterns behind the strength of persistent ties.
\newblock {\em EPJ Data Science}, 6(1):31, 2017.

\bibitem{paine16_differ_circad_phase_weekd_sleep}
Sarah-Jane Paine and Philippa~H. Gander.
\newblock Differences in circadian phase and weekday/weekend sleep patterns in
  a sample of middle-aged morning types and evening types.
\newblock {\em Chronobiology International}, 33(8):1009--1017, 2016.

\bibitem{panda02_circad_rhyth_from_flies_to_human}
Satchidananda Panda, John~B. Hogenesch, and Steve~A. Kay.
\newblock Circadian rhythms from flies to human.
\newblock {\em Nature}, 417(6886):329--335, 2002.

\bibitem{pinquart01_influen_lonel_older_adult}
Martin Pinquart and Silvia Sorensen.
\newblock Influences on loneliness in older adults: a meta-analysis.
\newblock {\em Basic and Applied Social Psychology}, 23(4):245--266, 2001.

\bibitem{roberts10_commun_social_networ}
Sam G.~B. Roberts and Robin~I.M. Dunbar.
\newblock Communication in social networks: Effects of kinship, network size,
  and emotional closeness.
\newblock {\em Personal Relationships}, 18(3):439--452, 2010.

\bibitem{roberts2011}
Sam~G.B. Roberts and Robin~I.M. Dunbar.
\newblock The costs of family and friends: an 18-month longitudinal study of
  relationship maintenance and decay.
\newblock {\em Evolution and Human Behavior}, 32(3):186--197, 2011.

\bibitem{saramaki2014persistence}
Jari Saram{\"a}ki, Elizabeth~A Leicht, Eduardo L{\'o}pez, Sam~GB Roberts, Felix
  Reed-Tsochas, and Robin~IM Dunbar.
\newblock Persistence of social signatures in human communication.
\newblock {\em Proceedings of the National Academy of Sciences},
  111(3):942--947, 2014.

\bibitem{sigmundova16_weekd_weeken_patter_physic_activ}
Dagmar Sigmundov{\'a}, Erik Sigmund, Petr Badura, Jana Vok{\'a}{\v{c}}ov{\'a},
  Lucie Trhl{\'i}kov{\'a}, and Jens Bucksch.
\newblock Weekday-weekend patterns of physical activity and screen time in
  parents and their pre-schoolers.
\newblock {\em BMC Public Health}, 16(1):898, 2016.

\bibitem{taillard21_sleep_timin_chron_social_jetlag}
Jacques Taillard, Patricia Sagaspe, Pierre Philip, and St{\'e}phanie Bioulac.
\newblock Sleep timing, chronotype and social jetlag: Impact on cognitive
  abilities and psychiatric disorders.
\newblock {\em Biochemical Pharmacology}, 191(nil):114438, 2021.

\bibitem{tamarit18_cognit_resour_alloc_deter_organ_person_networ}
Ignacio Tamarit, Jos{\'e}~A. Cuesta, Robin I.~M. Dunbar, and Angel S{\'a}nchez.
\newblock Cognitive resource allocation determines the organization of personal
  networks.
\newblock {\em Proceedings of the National Academy of Sciences},
  115(33):8316--8321, 2018.

\bibitem{tamarit22_beyon_dunbar_circl}
Ignacio Tamarit, Angel S{\'a}nchez, and Jos{\'e}~A. Cuesta.
\newblock Beyond dunbar circles: a continuous description of social
  relationships and resource allocation.
\newblock {\em Scientific Reports}, 12(1):2287, 2022.

\bibitem{williams18_inter_emotion_regul}
W.~Craig Williams, Sylvia~A. Morelli, Desmond~C. Ong, and Jamil Zaki.
\newblock Interpersonal emotion regulation: Implications for affiliation,
  perceived support, relationships, and well-being.
\newblock {\em Journal of Personality and Social Psychology}, 115(2):224--254,
  2018.

\bibitem{zhou2005organizational_social_groups}
W.-X. Zhou, D.~Sornette, R.~A. Hill, and R.~I.~M. Dunbar.
\newblock Discrete hierarchical organization of social group sizes.
\newblock {\em Proceedings of the Royal Society B: Biological Sciences},
  272(1561):439--444, 2005.

\end{thebibliography}
\bibliographystyle{plain}

\appendix
\subsection*{Availability of Data}
Data files and code to produce the analysis reported in this manuscript can be found in Vergara Hidd, Valentin et al. (2023), The Rhythms of Transient Relationships: Allocating Time between Weekdays and Weekends, Dryad, Dataset, \href{https://doi.org/10.5061/dryad.w6m905qv9}{https://doi.org/10.5061/dryad.w6m905qv9}

\subsection*{Competing interests}
The authors declare that they have no competing interests.

\subsection*{Funding}
No funding was received for this research.

\subsection*{Author's contributions}
  VVH, EL, MZ, and SR conceptualized and designed the analyses. VVH analyzed the data and created visualizations. VVH, EL, and SR wrote the manuscript. SR, SC, and BL provided access to the data.  All authors read, revised and approved the final manuscript.

\subsection*{Acknowledgements}
All authors thank Robin Dunbar for his discussion on the ideas of this paper, as well as comments on early drafts. SC and BL acknowledge The Mobile Territorial Lab (MTL), a joint initiative created by TIM - Telecom Italia, Fondazione Bruno Kessler, MIT Media Lab, and Telefonica Research. SC and BL also thank all the MTL study participants.

\end{document}